\documentclass[conference]{IEEEtran}
\ifCLASSINFOpdf
\else
\fi

\usepackage{fancyhdr}
\usepackage{mathtools}
\usepackage{graphicx}
\begin{document}
%
\title{Integration of Static and Dynamic Analysis for Malware Family Classification with Composite Neural Network}

\author{\IEEEauthorblockN{Yao Saint, Yen}
\IEEEauthorblockA{Institute of Information Science, Academia Sinica, Taiwan\\
yenyaosam@iis.sinica.edu.tw}
\and
\IEEEauthorblockN{Zhe Wei, Chen}
\IEEEauthorblockA{Institute of Information Science, Academia Sinica, Taiwan\\
zw062636@iis.sinica.edu.tw}
\and
\IEEEauthorblockN{Ying Ren, Guo}
\IEEEauthorblockA{Institute of Information Science, Academia Sinica, Taiwan\\
hitoshi@iis.sinica.edu.tw}
\and
\IEEEauthorblockN{Meng Chang, Chen}
\IEEEauthorblockA{Institute of Information Science, Academia Sinica, Taiwan\\
mcc@iis.sinica.edu.tw}}


%



\maketitle

\thispagestyle{fancy}
\lhead{} 
\chead{} 
\rhead{} 
\lfoot{} 
\cfoot{\thepage} 
\rfoot{} 
\renewcommand{\headrulewidth}{0pt} 
\renewcommand{\footrulewidth}{0pt} 
\pagestyle{fancy}
\cfoot{\thepage}

\begin{abstract}
Deep learning has been used in the research of malware analysis. Most classification methods use either static analysis features or dynamic analysis features for malware family classification, and rarely combine them as classification features and also no extra effort is spent integrating the two types of features. In this paper, we combine static and dynamic analysis features with deep neural networks for Windows malware classification. We develop several methods to generate static and dynamic analysis features to classify malware in different ways. Given these features, we conduct experiments with composite neural network, showing that the proposed approach performs best with an accuracy of 83.17\% on a total of 80 malware families with 4519 malware samples. Additionally, we show that using integrated features for malware family classification outperforms using static features or dynamic features alone. We show how static and dynamic features complement each other for malware classification.
\end{abstract}

\section{Introduction}
Malware is a threat to both individuals and industry; numerous malware programs are detected every day. Many malware programs share similar malicious behavior. Experts speculate that such malware may have the same origin, that is, later malware could be the result of modifications and mixtures of existing malware. In order to better manage and analyze malware, malware programs with similar attack patterns are considered members of the same family. Once a new malware program is discovered, a good practice is to first classify its family to help analysts to comprehend the malware and devise a mediation solution. If no family is identified, a new family is created for the program.

Most malware analysis uses either static analysis, which involves examining the malware program structure without executing it, generating features such as API calls, strings, header files, and call graphs; or dynamic analysis, which involves executing the malware program and collecting execution traces such as system and API call sequences and their associated parameters. Both static and dynamic analysis have their pros and cons: static analysis saves execution costs but may be vulnerable to complex code obfuscation. In dynamic analysis, in turn, running the malware program in a sandbox takes time and resources. 
Few studies combine both methods and optimize their effectiveness. In this study, we apply features from both static and dynamic analysis, using them to compensate for the shortcomings of each.

Deep learning is widely used in malware classification. Methods for experimentation include random forest, SVM (support vector machine), KNN (k-nearest neighbor), or other common machine learning methods; rarely are more complex modules used. In this study, we employ the composite neural network framework, which is composed of several pre-trained neural networks trained separately to solve part of the problem or the whole problem. Pre-trained neural network can be defined that a deep neural network, after training on a particular recognition task, can be applied on another domain. Yang et al.~\cite{yang2019theoretical} investigate the performance of composite neural networks, finding a guaranteed high probability that the performance of the composite network is better than any of its pre-trained components. Here, inspired by their findings, we use multiple pre-trained components with different topologies for pragmatic performance evaluation. 

In this paper, we consider a group of diverse features extracted from static and dynamic analyses for Windows malware classification. For static analysis, we first analyze the malware to collect the API call statistics. This is because code from family members generally has similar statistics; thus this feature is useful for classifying malware with non-sophisticated or non-major modifications. Then we extract the procedure calling relation and generate a call graph, represented by an adjacency matrix, after which we train a graph embedding model to transform a call graph into its embedding representation. Finally, for the adjacency matrix, we further collect most of the signal information in the call graph by using DCT (discrete cosine transform), which uses zigzag scanning to gather low-frequency components as one of the static features.
 
For dynamic analysis, we observe malware behavior by executing the malware program in a sandbox environment, recording API call sequences and their associated parameters as the program behavior in the form of a trace file. We calculate the frequency of each API call in this trace file, which we then use as a feature for family classification of the program. We then use hierarchical attention networks to process the trace file to generate an embedding representation, after which we build a co-occurrence matrix based on the trace file, and use a CNN (convolutional neural network) to extract its structural features. Finally, based on paragraph vectors, we use distributed representations of sentences and documents to embed the trace file as a trace embedding, constructing representations of variable-length input sequences.

This paper uses fusion neural network to build a malware classification model. The network is divided into early fusion neural network, late fusion neural network, and iterated fusion neural network. The early fusion neural network takes the features to be inputted into the classification model and then connects to a classifier; the late fusion neural network takes the results from the classification models, which can be seen as pre-trained models, and further connects to a classifier; finally, the iterated fusion neural network iteratively combines two pre-trained neural networks into one and continues the combination work until reaching the root of network. The results show that the proposed approach performs best, with a classification accuracy of 83.17\% on malware family classification. We show that combining both static and dynamic features to build a fusion neural network outperforms using only static features or dynamic features for classification. We show how static and dynamic features complement each other in malware classification, and then compare using simple API call sequences and API statement sequences to extract temporal characteristics, showing that most such approaches consider only the API call sequence and not additional information that could help the classifier. We show that considering the API call sequence and its associated parameters with the proper deep learning method outperforms using simple API call sequences with deep learning. These are the contributions of our work:
~
\begin{itemize}
  \item We propose classifying malware using several methods, including deep learning to extract features using static and dynamic analysis. To the best of our knowledge, this is the first paper to use static and dynamic analysis features with fusion neural network to classify malware on a Windows OS.
  \item We show that combining static and dynamic features yields classification performance superior to that using only static or dynamic features.
  \item We show how static and dynamic features
  complement each other during malware classification.
  \item We compare using simple API call sequences and using not only API call sequences but also their associated parameters; the results show that considering additional information in classification yields superior results.
\end{itemize}

\section{RELATED WORK}
In recent years, machine learning based malware classification approaches have demonstrated high efficiency and accuracy. Usually such approaches consist of feature extraction and a classification model. Researchers extract different features from malware using static or dynamic analysis and then use machine learning and standard deep learning approaches for classification.

Static analysis, is used to examine the program without executing it. De La Rosa et al.~\cite{de2018efficient} proposed a meta model that finds the simplest classifiers to characterize and assign malware into their corresponding families, their classifier features used three static analysis features to characterize a malware, include basic, byte and assembly code to do malware classification using their proposed meta-model. Hassen et al.~\cite{hassen2017malware} proposed a new feature based on control statement shingling with ordinary opcode n-gram based features to classify malware. Yan et al.~\cite{yan2018detecting} proposed an approach that using deep learning for malware detection which takes CNN and LSTM(long short term memory) to automatically learning features from the raw data, to get the file structure and code sequence patterns, then they further used stacking ensemble to combine networks’ results to optimize the detection accuracy. Kabanga et al.~\cite{kabanga2017malware} proposed a model that used machine learning’s convolution neural network to classify images extracted from malware binaries. Kim et al.~\cite{kim2017malware} proposed a transferred generative adversarial network (tGAN) for automatic classification and detection of the zero-day attack, and they to use tGAN, they convert malware codes to images, called malware images, used that as static analysis feature to train deep neural network. 
And in dynamic analysis, it is the testing and evaluation of a program by executing it and extract its program behavior, Pascanu et al.~\cite{pascanu2015malware} attempt to learn the language of malware of detecting these unknown threats, they used a recurrent model trained to predict next API call, and use the hidden state of the model as the fixed-length feature vector to train a classifier and that classifier made by MLP(multi-layer perceptron), one standard method of deep learning. Tobiyama et al.~\cite{tobiyama2016malware} proposed a malware process detection method using process behavior to detect whether a terminal is infected or not, they used two types of DNN, one is RNN to extract feature in program, the other one is CNN, it is used to do classification based on the aforementioned feature. Alsulami et al.~\cite{alsulami2018behavioral} contributed to behavioral malware classification using information gathered from Microsoft Windows Prefetch files, then they used Convolutional Recurrent Neural Networks to build their classification model. Agrawal et al.~\cite{agrawal2018robust} proposed a neural model that takes variable-length input sequences of system API calls, then they implement a version of Convoluted Partitioning of Long Sequences(CPoLS) which can capture sequences of any variable length to do malware classification. Smith et al.~\cite{smith2018dynamic} examined several machine learning techniques for detecting malware including random forest, deep learning techniques, and liquid state machines, with system API calls. Stokes et al.~\cite{stokes2017attack} implement and study several learning-based attacks and defenses for dynamic analysis, then applied deep learning method to do classification.
There also have methods that combine both static and dynamic analysis features to do malware detection or classification, such as Yuan et al.~\cite{yuan2016droiddetector} proposed to combine the static analysis features and features from dynamic analysis of Android apps and then using deep learning techniques to do malware detection. Chakraborty et al.~\cite{chakraborty2017ec2} used ensemble clustering and classification to classify malware, also show how to characterize different malware families by extracting family-specific features that distinguish one family from others, but this study did not emphasize the benefits of using both static and dynamic analysis to classify only mention when under certain circumstance with hybrid analysis will perform better. Onwuzurike et al.~\cite{onwuzurike2018family} used online system with a little modify to obtain both static and dynamic features, then apply random forest to classify malware, though it have phenomenon of granularity in static and dynamic analysis, it only consider random forest as classification method, didn't apply complex classification methods such as deep learning with more comprehensive view to do classification.

In building a deep learning based classification model, standard classic methods include SVM, random forest, KNN, CNN, RNN, and MLP. As deep learning has grown in popularity, many novel methods based on deep learning have been proposed that increase efficiency and accuracy. One example is ensemble learning, in which learning algorithms construct a set of classifiers and then classify new data points and use methods such as taking a vote on their predictions. Krawczyk et al.~\cite{krawczyk2018online} proposed a lightweight and flexible abstaining extension for online ensembles that allows excluding some classifiers from the voting process. And multi-view learning~\cite{xu2013survey}, it means by learning from multi-view data by considering the different features. Elkahky et al.~\cite{elkahky2015multi} proposed a novel deep learning approach extended from the Deep Structured Semantic Models (DSSM) to map users and items to a shared semantic space and recommend items that have maximum similarity with users in the mapped space.

To the best of our knowledge, most research on malware classification uses either static or dynamic analysis as features to classify malware directly; rarely do some extra effort on these features. Classification methods usually use standard machine learning or deep learning methods such as SVM, random forest, KNN, CNN, and RNN, and not complex methods such as composite neural networks. Hence, in our approach, we propose combining static and dynamic analysis features using fusion neural network for use in Windows malware classification.

\section{Proposed Framework}
\subsection{Static analysis}
We divide static analysis into two aspects: call graphs and API calls that might appear in a PE file. We further use two methods to extract from the call graph with two different feature sets that characterize its structure: call graph embedding with deep neural networks, and low-frequency regions from the call graph.

\subsubsection{API calls appear in PE}\hfill \break
The behavior of members of a given family of malware is similar, although malware authors may use techniques such as code obfuscation to avoid detection. We apply PEframe~\cite{Amato2019}, an open source tool for PE file static analysis, to extract static properties such as the API calls used in the malware. As there are 251 different API calls found in the malware corpus, a vector of 252 slots, in addition to \textsc{unknown} for all other API calls, is used for a one-hot encoding of the API call indicators, with 1 meaning the API is in the malware and 0  otherwise.

\subsubsection{DNN-based call graph embedding}
\begin{figure}[hbt!]
\centering
\includegraphics[width=0.5\textwidth]{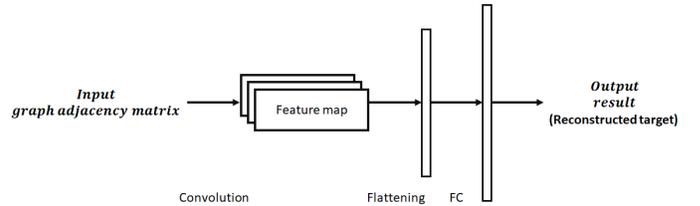}
\caption{CaFC (convolution and fully connect)} \label{CAFC}
\end{figure}

\hfill \break
The call graph (CG) is a representation of the function call relations among problem blocks; it is essential in static analysis. In this study, we obtain the CG using radare2 ~\cite{radare22019}, a complete framework for PE file reverse-engineering and analysis. We implemented a program to generate the adjacency matrix for the CG. Inspired by the autoencoder technique, which learns the characteristics of the training data in an unsupervised way, we propose the deep neural network shown in Fig.~\ref{CAFC}. The front stage is a CNN to catch significant patterns in the adjacency matrix, and the rear stage is an autoencoder composed of an encoder and decoder. The encoder maps the input into a lower dimensional internal representation, and decoder reverses the representation to reconstruct the original input. We term this model a CaFC (convoluted and fully connected) model. The encoder of the trained CaFC model is used as the call graph embedding model.

\subsubsection{Low-frequency CG regions}
\begin{figure}[hbt!]
\includegraphics[width=0.5\textwidth]{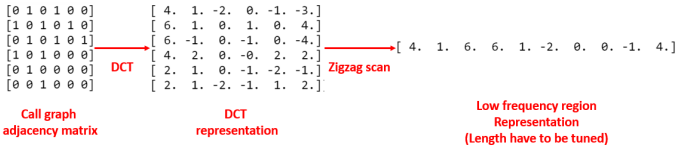}
\centering
\caption{Extraction of low-frequency CG regions. DCT is used to transform the call graph adjacency matrix into its DCT representation, after which we use zigzag scanning to scan low-frequency regions into a vector.} \label{LFR}
\end{figure}

\hfill \break
We use a non-training method to obtain the CG characteristics. Inspired by JPEG (joint photographic experts group), the most widely used image compression method on the Internet, we apply its most important parts. The first is DCT (discrete cosine transform), which compacts the energy inside the image; in typical applications, most of the signal information is concentrated on a few low-frequency DCT components. The second is zigzag scanning, by which low-frequency coefficients are grouped after using DCT. These two methods are used to extract low-frequency characteristics (here termed \emph{low-frequency regions}) from the CG. Fig.~\ref{LFR} shows how we extract such low-frequency CG regions.

\subsection{Dynamic analysis}
For dynamic analysis, we use Cuckoo sandbox~\cite{cuckoo2019}, the leading open-source automated malware analysis system. Cuckoo yields the following features:
\begin{itemize}
  \item API calls made by all malware processes
  \item Filenames created, deleted, and downloaded by malware
  \item Malware memory dumps
  \item Network traffic
\end{itemize}

In this work, to monitor PE file behavior, we gather the API calls that are made along with their parameters into the trace file, which we use as the foundation of further processes to extract malware characteristics. We propose a total of four methods to process the trace file: API call frequency, PV-based trace embedding, a DNN-based co-occurrence matrix, and an API statement sequence encoder.

\subsubsection{API call frequency}\hfill \break
We calculate the frequency of each API call in each file. As there are 286 different API calls in our trace file training data set, for each file, we first build a vector of length 287, one element of which is \textsc{unknown}. Each different API call is given an index. The API call frequencies are yielded by taking as the denominator the total number of API calls in the trace file, and taking as the numerator the number of times each API call appears in the trace file.

\subsubsection{PV-based trace embedding}\hfill \break
Use of language models with deep neural network has been demonstrated by ~\cite{pascanu2015malware}, ~\cite{athiwaratkun2017malware} on malware detection or classification tasks. In our work, the trace files are composed of a series of API call sequences and their parameters. In this section, we take a series of API calls as a document, and introduce a paragraph vector (PV) by which to embed our trace file. For the paragraph vector we make use of an unsupervised learning algorithm that learns fixed-length feature representations from variable-length text segments ~\cite{le2014distributed}. It learns paragraph and document embedding via the distributed memory and distributed bag-of-words models, both of which are implemented using a deep neural network. After training the paragraph vector model, we use the model to embed our trace into a trace embedding (here termed a \emph{PV-based trace embedding}).

\subsubsection{Co-occurrence matrix with DNN}
\begin{figure}[hbt!]
\includegraphics[width=0.5\textwidth]{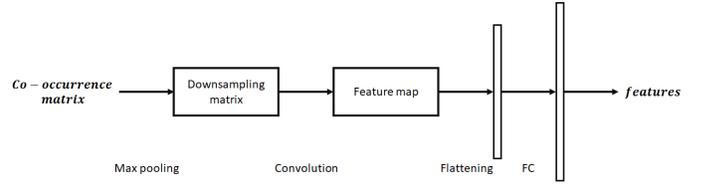}
\centering
\caption{Model to extract characteristics from co-occurrence matrix. The model uses max pooling to downsample the co-occurrence matrix, convolution to extract spatial features, flattening to represent the data in one dimension, and a fully connected layer to generate features from the co-occurrence matrix.} \label{CoDNN}
\end{figure}

\hfill \break
API call co-occurrences suggest that the API calls are closely related and could constitute certain action(s) which can be used to classify the malware family, as different malware families have different actions. For this reason, we use the co-occurrence matrix, which is composed of the frequency of term co-occurrences in a given order. We first build a co-occurrence matrix based on the trace file, after which we build a DNN model to extract the characteristics of the co-occurrence matrix. As shown in fig.~\ref{CoDNN}, this model is composed of four layers: a max pooling layer for downsampling, a convolution layer to extract significant events within the matrix, a flatten layer to turn data into a one-dimensional array, and a fully connected layer to extract features from the previous layer. The training target of this model is malware family classification. After training this model, we use it as a pre-trained model.

\subsubsection{API statement sequence encoder}
\begin{figure}[hbt!]
\includegraphics[width=0.45\textwidth]{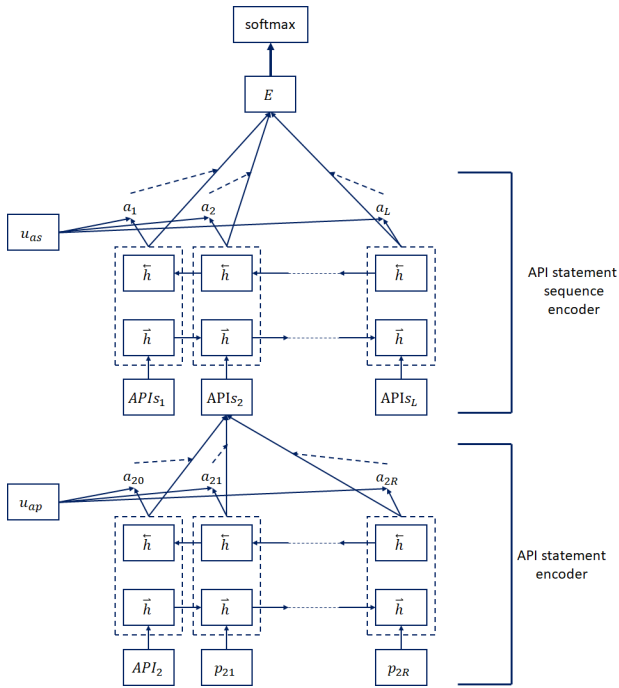}
\centering
\caption{API statement sequence model. Following Yang's~\cite{yang2016hierarchical}  approach, the training target is malware classification. It is divided into an API statement encoder (SE) and an API statement sequence encoder (SSE). After training, we use the API SE to extract characteristics from API statement, then using the API SSE to extract the temporal characteristics from API statement sequence.} \label{APISSE}
\end{figure}

\hfill \break
The trace file consists of an API statement sequence, that is, a sequence of API calls and their parameters. Following Yang's~\cite{yang2016hierarchical} approach, we propose the model in Fig.~\ref{APISSE}, termed an API statement sequence model. We train the model as a pre-trained model to encode the trace file's temporal series features to a fixed-length numerical vector as a representation of the malware program's behavior. Trained to classify malware, the model can be divided into an API statement encoder (SE) and an API statement sequence encoder (SSE). The API SE calculates the attention weight~\cite{vaswani2017attention} of each API call and its parameters using a bi-directional LSTM and a context weight matrix. The inner product is used to calculate the output from the bi-directional LSTM and the context weights to generate the attention weights. The API SE uses the following notation:
\begin{itemize}
  \item {$\mathit{API}_{l}$:} The $l$-th API call.
  \item {$p_{lr}$:} Parameter $r$ of API call $l$.
  \item {$\overrightarrow{h}$:} LSTM hidden state from beginning to end.
  \item {$\overleftarrow{h}$:} LSTM hidden state from end to beginning.
  \item {$u_{ap}$:} The context weight matrix of the API calls and their parameters.
  \item {$a_{lr}$:} The attention weight of API call $l$ and parameter $r$.  
\end{itemize}

Given the attention weight of each API call and its parameters, we build an API statement sequence encoder to encode each API statement in the trace file and generate attention weights for each API statement. The architecture of the API SSE depends on how many API statements are in the trace file. For each API statement, we calculate its API statement embedding using the attention weights from the API SE and the output from the bi-directional LSTM, using the inner product as the API statement embedding. Then, given every API statement embedding, we use the inner product of the bi-directional LSTM and another context weight matrix to generate the attention weight, after which we calculate the API call sequence embedding as the inner product of the API statement attention weight and the output of the bi-directional LSTM. Given the API call sequence embedding, we use a fully connected layer with softmax activations to classify the malware. This constitutes the API call sequence model, the notation for which is as follows.
\begin{itemize}
  \item {$\mathit{APIs}_{l}$:} The $l$-th API statement in the trace file.
  \item {$\overrightarrow{h}$:} LSTM hidden state from beginning to end.
  \item {$\overleftarrow{h}$:} LSTM hidden state from end to beginning.
  \item {$u_{as}$:} The context weight matrix of API statement.
  \item {$a_{l}$:} The attention weight of API statement $l$.
\end{itemize}
After training, we use the API SE and API SSE as tools to extract API call sequence embedding.

\section{Experimental results}
Here we describe the data collection and labeling, the setting of tunable features, and the results from different composite neural networks. We compare static features, dynamic features, and integrated features, and also compare the API statement sequence encoder with the API call sequence encoder.

\subsection{Data Collection and Labeling}

As shown in fig.~\ref{tab19}, we used 80 malware families with a total of 4519 malware samples collected by the National Center for High-Performance Computing (NCHC)~\cite{NCHC2019}, Taiwan, the distribution has shown our data are most balanced. We used traces provided by Cuckoo Sandbox and record only system API calls in our scenario. As there is no ground truth for malware families, Hsiao et al.~\cite{hsiao2013cooperative} ~\cite{hsiao2017virtual} leverage peer voting on labels from popular intrusion detection systems (IDSs) to determine malware families, which is shown to be better than most IDS labels; we adopted their malware family classification as the family label in our work, there are some notable malware family in our data set, such as malware family {\sl fakeav} with 188 samples, {\sl ramnit} with 90 samples, etc. In all reported experiments, we divided our data into 3661 training data, 407 validation data, and 451 testing data, and applied ten-fold cross validation to build our models. 

\begin{table}[hbt!]
\centering
\caption{Malware Family Samples Quantity Distribution}\label{tab19}
 \begin{tabular}{|c | c |} 
 \hline Distribution of malware samples & $\#$ of family belong to this distribution  \\
 \hline 0 $\sim$ 50    & 49  \\
 \hline 51 $\sim$ 100  & 19  \\
 \hline 101 $\sim$ 150 &  7  \\
 \hline 151 $\sim$ 200 &  2  \\
 \hline 201 $\sim$ 250 &  1  \\
 \hline 251 $\sim$ 300 &  1  \\
 \hline 301 $\sim$ 350 &  1  \\
 \hline
\end{tabular}
\end{table}


\subsection{Setting Tunable Features}
Some of our features are tunable and some are not tunable. ``Tunable'' means we can tune the length of the feature vector to determine which length best fits our classification task. For non-tunable features, the feature vector's length is fixed. To determine the length of a tunable feature, we trained a malware classification model---in this case a single fully connected layer using softmax activations---using various feature lengths, and used the length which performed best as the final feature length. The tunable features in our work are low-frequency CG regions, DNN-based CG embedding, PV-based trace embedding, the DNN-based co-occurrence matrix, and the API statement sequence encoder. We review these separately.

\subsubsection{DNN-based call graph embedding}\hfill \break
In this part, the tunable feature is the number of filters in the convolution layer. We determined the number of feature maps by deciding the number of kernels. In a CNN, kernels capture the different features of the call graph. We experimented with 3 kernels to 7 kernels to determine which number of kernels performs better on malware family classification. As shown in the table~\ref{tab1}, choosing 4 filters yielded the best performance.

\begin{table}[hbt!]
\centering
\caption{Accuracy with kernels of DNN-based call graph embedding}\label{tab1}
 \begin{tabular}{|c | c |} 
 \hline Number of kernels & Accuracy\\
 \hline
 3 & 31.31\%\\ 
 \hline
 4 & 31.42\%\\
 \hline
 5 & 31.22\%\\
 \hline
 6 & 31.24\%\\
 \hline
 7 & 30.93\%\\
 \hline
\end{tabular}
\end{table}

\subsubsection{Low-frequency Call Graph regions}\hfill \break
For low-frequency CG regions, we tuned the scanning length in zigzag scanning, as shown in Fig.~\ref{LFR}. The scanning length determines how many low-frequency region we want to use. We experimented with lengths from 150 to 350 for malware classification. As shown in the table~\ref{tab2}, choosing 350 as the scanning length yielded the best performance.

\begin{table}[hbt!]
\centering
\caption{Accuracy with zigzag scanning lengths of Low-frequency CG regions}\label{tab2}
 \begin{tabular}{|c | c |} 
 \hline Zigzag scanning length & Accuracy\\
 \hline
 150 & 31.12\%\\ 
 \hline
 200 & 31.13\%\\
 \hline
 250 & 31.18\%\\
 \hline
 300 & 30.75\%\\
 \hline
 350 & 31.26\%\\
 \hline
 400 & 31.06\%\\
 \hline
\end{tabular}
\end{table}

\subsubsection{PV-based trace embedding}\hfill \break
For PV-based trace embedding, we chose the size of the embedding while training the paragraph vector. We experimented with sizes from 100 to 500. and the result has shown in table~\ref{tab3}. An embedding size of 400 performed best.

\begin{table}[hbt!]
\centering
\caption{Accuracy with embedding sizes of PV-based trace embedding}\label{tab3}
 \begin{tabular}{|c | c |} 
 \hline Embedding size & Accuracy\\
 \hline
 100 & 74.63\%\\ 
 \hline
 200 & 75.74\%\\
 \hline
 300 & 75.21\%\\
 \hline
 400 & 76.01\%\\
 \hline
 500 & 75.39\%\\
 \hline
\end{tabular}
\end{table}

\subsubsection{DNN-based co-occurrence matrix}\hfill \break
For the DNN-based co-occurrence matrix, as shown in fig~\ref{CoDNN}, after turning the trace file into a co-occurrence matrix, we use it to train a classification model as a pre-trained model for malware classification. For this we tuned the pooling size of the model's max pooling layer. We tested sizes from 8 to 32. As shown in the table~\ref{tab4}, a pooling size of 4 performed best.

\begin{table}[hbt!]
\centering
\caption{Result of Accuracy with pooling sizes of Co-occurrence matrix}\label{tab4}
 \begin{tabular}{|c | c |} 
 \hline pooling size & Accuracy\\
 \hline
 4 & 59.43\%\\ 
 \hline
 8 & 61.70\%\\ 
 \hline
 16 & 56.48\%\\
 \hline
 32 & 44.63\%\\
 \hline
\end{tabular}
\end{table}

\subsubsection{API statement sequence encoder}\hfill \break
For this we tuned the number of API statements in the trace file, which we use to
train the API statement sequence model for malware classification.
We experimented with API statement sequence lengths from 100 to 400. As
shown in the table~\ref{tab5}, although a length of 300 yielded the best performance,  
the execution time for 200 was much shorter than that of 300; 
thus we chose a sequence length of 200, even though it was slightly outperformed 
by 300.

\begin{table}[hbt!]
\centering
\caption{Accuracy with sequence lengths in API statements sequence  encoder.}\label{tab5}
 \begin{tabular}{|c | c |} 
 \hline Sequence length & Accuracy\\
 \hline
 100 & 65.70\%\\ 
 \hline
 200 & 67.92\%\\ 
 \hline
 300 & 68.08\%\\
 \hline
 400 & 64.74\%\\
 \hline
\end{tabular}
\end{table}
\subsection{Experimental Results with Different Composite Neural Network Models}
After tuning features, we build the classification model. There are many ways to accomplish this; in this paper, we use composite neural networks, which consist of early fusion model and late fusion model with different topology. And in the training phase, we tune the hyperparameters of every model shown in the table~\ref{tab9}. These hyperparameters control the regularization used for each DNN layer, the dropout rate, whether batch normalization is used, and how many hidden layers to use and how many neurons in each layer, we also use ten-fold cross validation in every experiments. In addition, we further take SVM, random forests, and late fusion ensemble learning as baseline methods for comparison with the proposed models.
All the aforementioned methods we test with static features, dynamic features and integrated features, to see whether integrated features will help the classification task mostly.

\begin{table}[hbt!]
\centering
\caption{The tunable hyper-parameter.}\label{tab9}
 \begin{tabular}{|c | c |} 
 \hline  Tuning parameter & Tuning ranges \\
 \hline
 Activation of each layer & {relu, sigmoid, tanh, softmax, linear}\\ 
 \hline
 Regulization & {0 \char`\~\hspace{0.05cm} 0.001}\\ 
 \hline
 Dropout rate of each layer & {0 \char`\~\hspace{0.05cm} 0.5}\\
 \hline
 Batch normalization layer & {yes, no}\\ 
 \hline
 Weight setting & {fixed, trainable}\\ 
 \hline
\end{tabular}
\end{table}

\subsubsection{SVM and Random Forest}\hfill \break
SVM is a learning model that analyzes data used for classification and regression analysis. The result has shown in table~\ref{tab11}, we can see though dynamic features based SVM model slightly lower than integrated features based SVM model, integrated features perform the best accuracy with SVM model.

\begin{table}[hbt!]
\centering
\caption{SVM Classification Result}\label{tab11}
 \begin{tabular}{|c | c | c |} 
 \hline Features & Accuracy & Top-3 accuracy \\
 \hline Integrated & 73.29\% & 78.60\% \\
 \hline Static & 68.73\% & 73.93\% \\
 \hline Dynamic & 73.09\% & 78.40\% \\
 \hline
\end{tabular}
\end{table}

The random forest algorithm is a classic method for classification, regression, and other tasks that operates by constructing a multitude of decision trees. The result has shown in table~\ref{tab13}, just like SVM model, integrated features performs better.

\begin{table}[hbt!]
\centering
\caption{Random Forest Classification Result}\label{tab13}
 \begin{tabular}{|c | c | c |} 
 \hline Features & Accuracy & Top-3 accuracy \\
 \hline Integrated & 77.45\% & 82.76\% \\
 \hline Static & 67.65\% & 75.73\% \\
 \hline Dynamic & 77.50\% & 79.40\% \\
 \hline
\end{tabular}
\end{table}

\subsubsection{Late Fusion Ensemble Learning}\hfill \break
We use late fusion ensemble learning with trained weights and fixed weights. Firstly, we use the proposed seven features to train seven pre-trained models, the goal of which is malware classification. The length of the output of each is equal to the number of malware families, and represents the probability of each malware family, their pre-trained models architectures are all the same but with separate training. The last layer is the fully connected 80-node softmax layer. As the result shown in the table~\ref{tab7}.

\begin{table}[hbt!]
\centering
\caption{Malware classification models trained by each feature.}\label{tab7}
 \begin{tabular}{|c | c | c |} 
 \hline Feature name & Accuracy\\
 \hline
 API calls appear in PE & 63.75\%\\ 
 \hline
 Call graph embedding with DNN & 31.42\%\\
 \hline
 Low frequency region from call graph & 31.26\%\\
 \hline
 API call frequency & 72.18\%\\
 \hline
 PV-based trace embedding & 76.01\%\\ 
 \hline
 Co-occurrence matrix with DNN & 59.43\%\\ 
 \hline
 API statement sequence encoder & 67.92\%\\ 
 \hline
\end{tabular}
\end{table}

\begin{figure}[hbt!]
\includegraphics[width=0.5\textwidth]{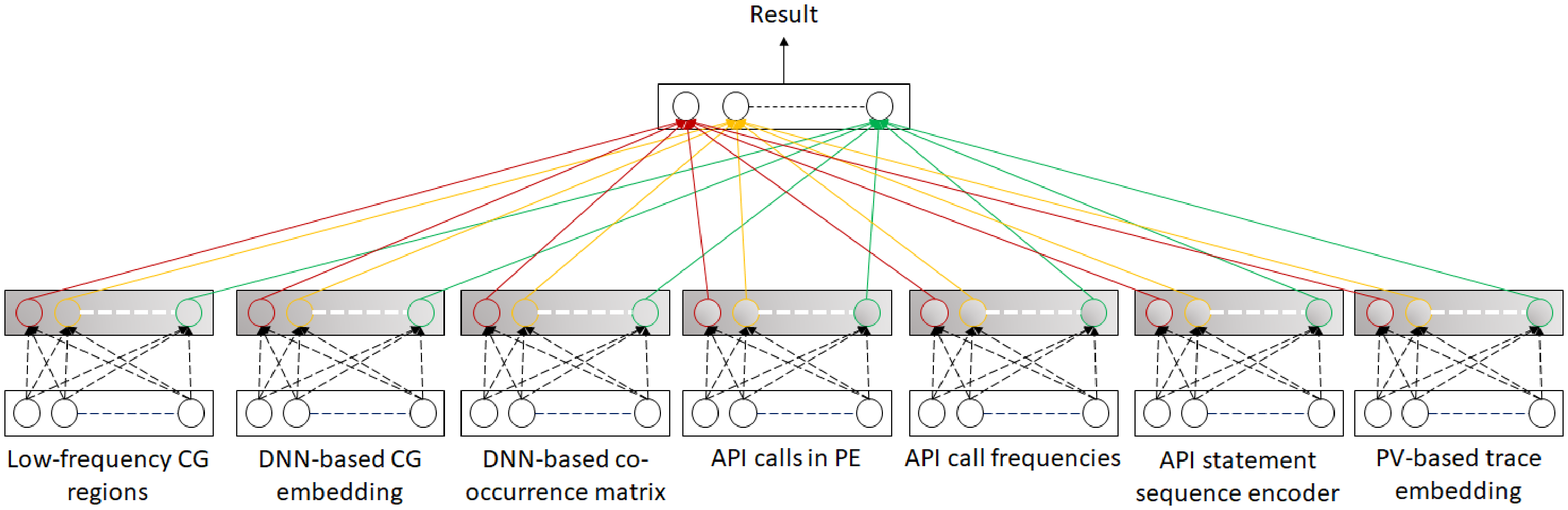}
\centering
\caption{Late Fusion Ensemble Learning structure with integrated features.} \label{Ensemble_structure}
\end{figure}

After training, we take every pre-trained models' (total 7 pre-trained models) single output (1 of 80) to concatenate them into one vector, which we then use to train a one-versus-all classifier which indicates whether it belongs to the family or not. Thus we train 80 classifiers, one for each family in our dataset. From these classifiers we choose the highest probability as the predicted label. We use both fixed and trainable weights in the classifier. 
For fixed weights, we set all weights equally, and use that weight for malware classification. Trainable weights are initialized using a random uniform distribution and are then updated during training.
Fig~\ref{Ensemble_structure} is used for late fusion ensemble learning using static and dynamic features to do classification. Late Fusion Ensemble Learning structure with integrated features. we take the single output (1 of 80) of every pre-trained models (total 7 pre-trained models) to concatenate them into one vector to train a one-versus-all classifier which indicates whether it belongs to the family or not. The multicolored line represent the individual output of each model, shaded gray blocks represent pre-trained layers and dashed arrows represent fixed weights. To further compare between integrated features, static features and dynamic features on classification task, we conducts the same experiments using static features and dynamic features separately.

The results shown in table~\ref{tab14}, dynamic features based late fusion ensemble learning almost be on an equal footing with integrated features based late fusion ensemble learning, though static features based late fusion ensemble learning is a lot lower than dynamic features based, but when integrated both of them, will perform better.

\begin{table}[hbt!]
\centering
\caption{Late Fusion Ensemble Learning Classification Results}\label{tab14}
 \begin{tabular}{|c | c | c |} 
 \hline Features & Acc & Top-3 Acc \\
 \hline Integrated with trainable weight & 74.70\% & 89.84\% \\
 \hline Static with trainable weight & 64.92\% & 82.84\% \\
 \hline Dynamic with trainable weight & 74.05\% & 88.25\% \\
 \hline Integrated with fixed weight & 72.79\% & 88.63\% \\
 \hline Static with fixed weight & 27.76\% & 51.97\% \\
 \hline Dynamic with fixed weight & 71.04\% & 89.14\% \\
 \hline
\end{tabular}
\end{table}

\subsubsection{Early fusion neural network}
\begin{figure}[hbt!]
\includegraphics[width=0.5\textwidth]{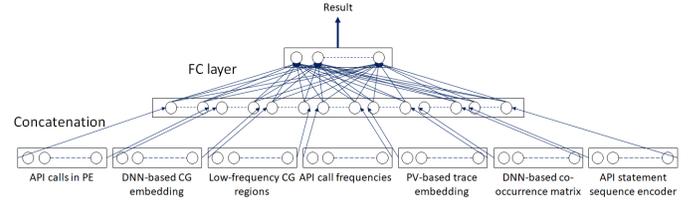}
\centering
\caption{Early fusion neural network 1 with integrated features.} 
\label{EarlyFusion}
\end{figure}

\begin{figure}[hbt!]
\includegraphics[width=0.5\textwidth]{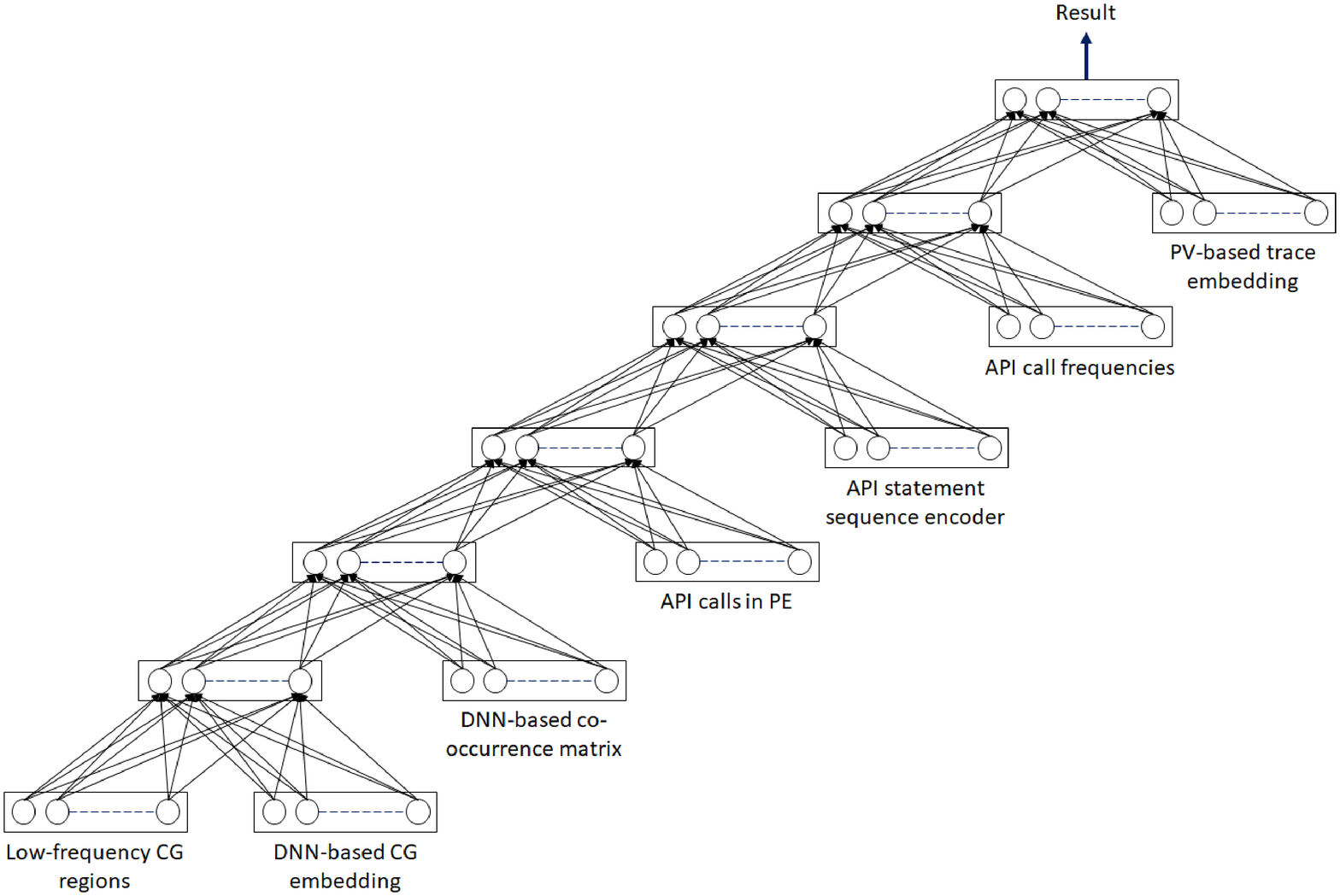}
\centering
\caption{Early fusion neural network 2 with integrated features. 
} 
\label{EarlyFusion2}
\end{figure}

\hfill \break
In the early fusion neural network, we propose two late fusion neural network models shown in fig~\ref{EarlyFusion} and fig~\ref{EarlyFusion2}. 
In early fusion neural network 1, the features are concatenated as a vector paaed through a fully connected layer for malware classification. After concatenate all these features, we use a softmax layer with shape 80 to classify our malware. The compared results of integrated features, dynamic features and static features are shown in table~\ref{tab15} that the integrated model performs the best result in our experiment.

\begin{table}[hbt!]
\centering
\caption{Early Fusion Neural Network 1 Classification Result}\label{tab15}
 \begin{tabular}{|c | c | c |} 
 \hline Features & Acc & Top-3 Acc \\
 \hline Integrated with trainable weight & 83.17\% & 90.53\% \\
 \hline Static with trainable weight & 66.63\% & 83.48\% \\
 \hline Dynamic with trainable weight & 81.81\% & 89.80\% \\
 \hline
\end{tabular}
\end{table}

In fig~\ref{EarlyFusion2}, our design idea is we choose the two pre-trained models with the poorest accuracy: the models using the DNN-based CG embedding and the low-frequency CG regions. The order of the model results from bottom left to upper right is in ascending accuracy: low-frequency CG regions (31.26\%), DNN-based CG embedding (31.42\%), DNN-based co-occurrence matrix (59.43\%), API calls in PE (63.75\%), API statement sequence encoder (67.92\%), API call frequencies (dynamic side, 72.18\%), and the PV-based trace embedding (76.01\%). Solid arrows represent trainable weights. We concatenate the output of these two models to concatenate a FC layer, then we concatenate the output from the first step with the output from the next worst model's output---in this case, that for the API statement sequence encoder---to connect another FC layer. Continuing in this fashion, we build early fusion neural network 2. The compared result of integrated features, dynamic features and static features are shown in table~\ref{tab16}, in this model structure, we can see dynamic based structure beats the integrated based structure. We believe the reason is in integrated feature based structure, we have too many pre-trained layers, and layer by layer stack it up, the result may have been formed earlier step, so will cause the poor result.

\begin{table}[hbt!]
\centering
\caption{Early Fusion Neural Network 2 Classification Result}\label{tab16}
 \begin{tabular}{|c | c | c |} 
 \hline Features & Acc & Top-3 Acc \\
 \hline Integrated with trainable weight & 80.71\% & 88.69\% \\
 \hline Static with trainable weight & 66.08\% & 81.81\% \\
 \hline Dynamic with trainable weight & 81.15\% & 87.58\% \\
 \hline
\end{tabular}
\end{table}

\subsubsection{Late fusion neural network}
\begin{figure}[hbt!]
\includegraphics[width=0.5\textwidth]{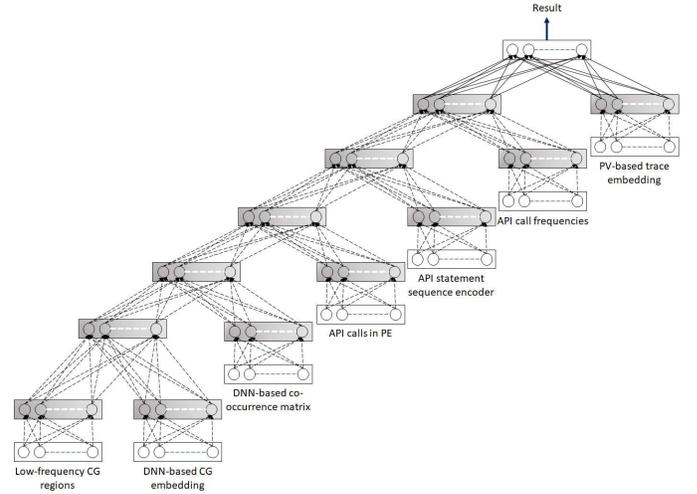}
\centering
\caption{Late fusion neural network 1 with integrated feature. 
} \label{LateFusion1}
\end{figure}

\begin{figure}[hbt!]
\includegraphics[width=0.5\textwidth]{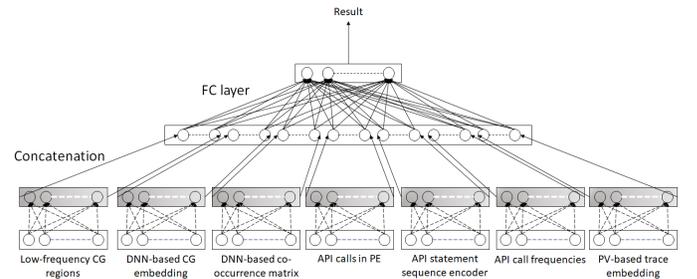}
\centering
\caption{Late fusion neural network 2 with integrated features.} \label{LateFusion2}
\end{figure}

\hfill \break
In the late fusion neural network, just like late fusion ensemble learning, we use the seven features to train seven models separately. The training goal of every model is malware classification, and their architectures are all the same but with separate training. The last layer is the fully connected 80-node softmax layer. As the result shown in the table~\ref{tab7}, we now have seven pre-trained models, the results of which we use to classify malware. We thus propose the two late fusion neural network models shown in
figures~\ref{LateFusion1} and \ref{LateFusion2}. 

In fig.~\ref{LateFusion1}, we choose the two pre-trained models with the poorest accuracy: the models using the DNN-based CG embedding and the low-frequency CG regions. We concatenate the output of these two pre-trained models to build a new pre-trained model, the target of which is also malware classification. Then we concatenate the output from the first step with the output from the next worst pre-trained model---in this case, that for the API statement sequence encoder---to build another new pre-trained model. Continuing in this fashion, we build late fusion neural network 1. We also compare the performance with static features and dynamic features, as shown in table~\ref{tab17}.

\begin{table}[hbt!]
\centering
\caption{Late Fusion Neural Network 1 Classification Result}\label{tab17}
 \begin{tabular}{|c | c | c |} 
 \hline Features & Acc & Top-3 Acc \\
 \hline Integrated with trainable weight & 74.50\% & 85.81\% \\
 \hline Static with trainable weight & 32.15\% & 57.70\% \\
 \hline Dynamic with trainable weight & 46.34\% & 54.10\% \\
 \hline
\end{tabular}
\end{table}

For late fusion neural network 2, as shown in fig~\ref{LateFusion2}, 
as in fig.~\ref{EarlyFusion}, we concatenate the outputs of the pre-trained models to a vector which is fed to a classifier with a fully connected layer and softmax activations for malware classification. The difference between this figure and fig.~\ref{EarlyFusion} is that here, we use the results of the pre-trained models, which are trained by the tunable and non-tunable features, and then concatenate these results as a vector of length 560 (80 families $\times$ 7 models) which is fed to a fully connected layer for classification, And for the aspect of compare static, dynamic and integrated features, and result has shown in table~\ref{tab18}, we can see when used integrated features based model structure performs best result.

\begin{table}[hbt!]
\centering
\caption{Late Fusion Neural Network 2 Classification Result}\label{tab18}
 \begin{tabular}{|c | c | c |} 
 \hline Features & Acc & Top-3 Acc \\
 \hline Integrated with trainable weight & 77.89\% & 89.80\% \\
 \hline Static with trainable weight & 69.18\% & 81.59\% \\
 \hline Dynamic with trainable weight & 77.16\% & 88.92\% \\
 \hline
\end{tabular}
\end{table}

\subsubsection{Iterative fusion neural network}
\begin{figure}[hbt!]
\includegraphics[width=0.5\textwidth]{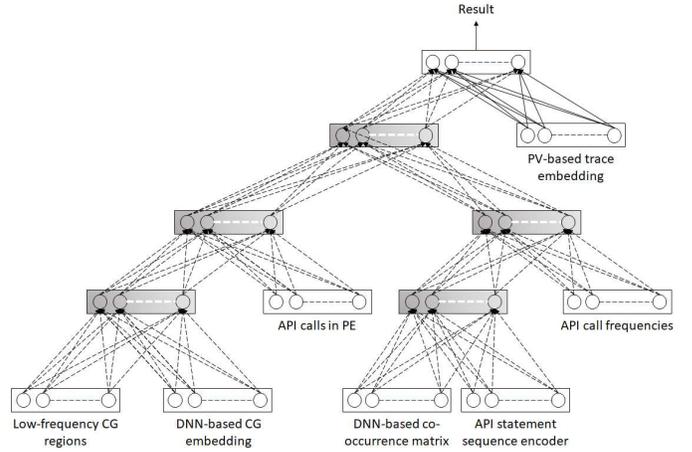}
\centering
\caption{Iterated fusion neural network 1.}
\label{CompositeFusion1}
\end{figure}

\begin{figure}[hbt!]
\includegraphics[width=0.5\textwidth]{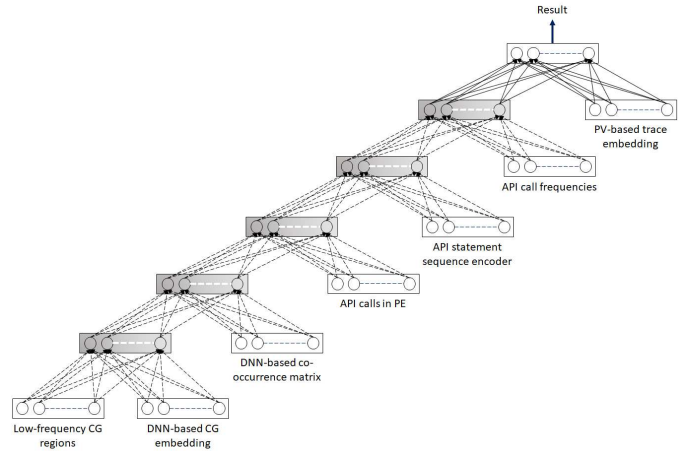}
\centering
\caption{Iterated fusion neural network 2. Features yielding higher accuracy 
are located closer to the classifier, so that features yielding lower accuracies 
are forced to pass through more models to aid the final classifier.} \label{CompositeFusion2}
\end{figure}

\hfill \break
As its name implies, iterative fusion neural network iteratively combines two pre-trained neural networks into one and continues the combination work until reaching the root of network. We
propose the models shown in fig~\ref{CompositeFusion1} and ~\ref{CompositeFusion2}.
In fig.~\ref{CompositeFusion1}, it is a tree structure model with the final 
layer as the root of tree. On the right half of the tree we choose the trace 
embedding feature because using this feature to train the malware classifier
yields the best performance; thus we want it to be closer to the root. 
The left half of the tree is also a tree structure: a pre-trained model 
for malware classification. Likewise for this tree's left and right halves: these are also pre-trained models, the difference being that one is mainly static analysis and the other dynamic analysis. The left half uses features from the CG embedding and low-frequency regions in the pre-trained model, and after training, concatenates the output of this pre-trained model with the API call features to train another malware classifier. In the right half of this tree, the co-occurrence matrix and the API statement sequence encoder are used to train a pre-trained model, the output of which is combined with the API call frequencies to train this tree structure model. Training the left half of the iterated fusion neural network in this way allows us to investigate the performance when static features and dynamic features are trained separately and then combined. And the result has shown in tab~\ref{tab20}, we can see integrated features performs best.

\begin{table}[hbt!]
\centering
\caption{Iterative fusion neural network 1 Classification Result}\label{tab20}
 \begin{tabular}{|c | c | c |} 
 \hline Features & Acc & Top-3 Acc \\
 \hline Integrated with trainable weight & 77.60\% & 85.14\% \\
 \hline Static with trainable weight & 32.15\% & 57.70\% \\
 \hline Dynamic with trainable weight & 46.34\% & 54.10\% \\
 \hline
\end{tabular}
\end{table}

Fig.~\ref{CompositeFusion2} is similar to Fig.~\ref{LateFusion1}, the
difference being that in fig.~\ref{LateFusion1}, every tunable and non-tunable feature
is used to train a pre-trained model, the outputs are which are then used in
further operations. In fig.~\ref{CompositeFusion2}, however, we
use the features directly. As in fig.~\ref{LateFusion1}, the order from
bottom left to upper right is in ascending accuracy. And the results has shown in tab~\ref{tab21}, as we can see the static and dynamic results are same with tab~\ref{tab20} is because when they build static and dynamic based structure will share the same structure, so the results will be the same.

\begin{table}[hbt!]
\centering
\caption{Iterative fusion neural network 2 Classification Result}\label{tab21}
 \begin{tabular}{|c | c | c |} 
 \hline Features & Acc & Top-3 Acc \\
 \hline Integrated with trainable weight & 79.37\% & 87.36\% \\
 \hline Static with trainable weight & 32.15\% & 57.70\% \\
 \hline Dynamic with trainable weight & 46.34\% & 54.10\% \\
 \hline
\end{tabular}
\end{table}

In the table~\ref{tab10}, we put every based on integrated features models' accuracy together, the best result has shown using early fusion neural network 1 yields a classification accuracy of 83.17\% and a top-3 accuracy of 90.53\%, outperforming the baseline methods. Also, most fusion neural network models perform better than the baseline methods, but the late fusion neural network 1 is outperformed by one of the baseline methods. This was due to degradation caused by an excessive number of layers and too many trainable weights. Thus when building a complex neural network such as those in our experiment, the neural network structure should closely fit the scenario.

\begin{table}[hbt!]
\centering
\caption{Comparison of results between each classification methods.}\label{tab10}
 \begin{tabular}{|c | c | c |} 
 \hline  Classification model & Acc & Top-3 Acc  \\
 \hline
 Early fusion neural network 1 & 83.17\% & 90.53\%\\
 \hline
 Early fusion neural network 2 & 80.71\% & 88.69\%\\
 \hline
 Late fusion neural network 1 & 74.50\% & 85.80\%\\ 
 \hline
 Late fusion neural network 2 & 77.89\% & 89.80\%\\ 
 \hline
 Iterated fusion neural network 1 & 77.60\% & 85.14\%\\ 
 \hline
 Iterated fusion neural network 2 & 79.37\% & 87.36\%\\ 
 \hline
 Ensemble learning with trainable weight & 75.60\% & 89.57\%\\ 
 \hline
 Ensemble learning with fixed weight & 72.79\% & 88.62\%\\ 
 \hline
 SVM & 73.29\% & 78.60\%\\ 
 \hline
 Random forest & 77.45\% & 82.76\%\\ 
 \hline
\end{tabular}
\end{table}

\subsection{How Static and Dynamic Features Work Together}

Next, we show how static and dynamic features complement each other.
We consider the following cases:

\begin{enumerate}
  \item Failure to classify malware with only dynamic features, or only static features.
  \item Success classifying malware with static feature.   
  \item Success classifying malware with integrated features.
\end{enumerate}

\begin{figure}[hbt!]
\includegraphics[width=0.5\textwidth]{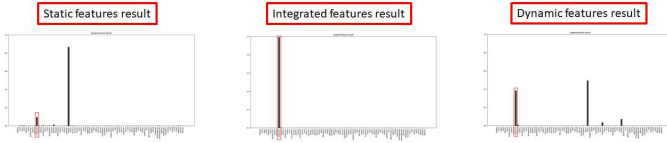}
\centering
\caption{Case 1 of static and dynamic features working together. Vertical
axis is probability and horizontal axis is each malware
family. Static features err egregiously, assigning a low probability 
to `softplus' and predicting the wrong malware family.
Dynamic features also mispredict the family. Integrated
features eliminate probabilities for other families and predict the right
one.} \label{case1}
\end{figure}

The first case is from the `softplus' malware family. Although this was not classified as malware with either static or dynamic features, it was successfully classified as malware with integrated features, as shown in fig.~\ref{case1}.

\begin{figure}[hbt!]
\includegraphics[width=0.5\textwidth]{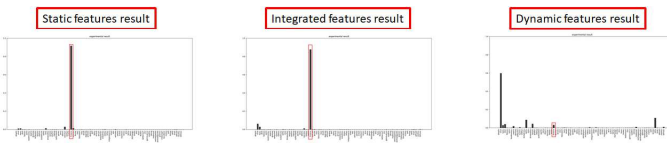}
\centering
\caption{Case 2 of static and dynamic features working together. 
Static features predict the family successfully, but dynamic features assign 
a low probability to `firseria'. Integrated
features preserve the good result from static features, assigning a high 
probability to `firseria'.} \label{case2}
\end{figure}

Case 2 is from the `firseria' malware family. As shown in fig.~\ref{case2},
this malware program is successfully
classified using static features, is misclassified using dynamic features, 
and is successfully classified using integrated features.

\begin{figure}[hbt!]
\includegraphics[width=0.5\textwidth]{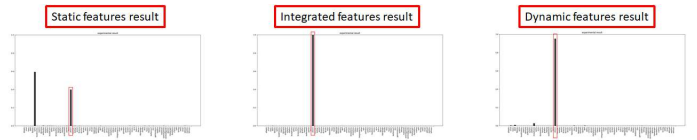}
\centering
\caption{Case 3 of static and dynamic features working together.
As in case 2, dynamic features predict the family successfully. For the
static features result, although it predicts the wrong family, it assigns a high
probability to `elkern'. Integrated features result correct the static features' 
erroneous probabilities and retain the good probabilities. } \label{case3}
\end{figure}

In case 3, for the `elkern' malware family, the malware program 
is misclassified using static features and successfully classified 
using dynamic features as well as integrated
features, as shown in fig.~\ref{case3}.

We thus observe that this methodology makes complementary use of most
static and dynamic features for malware family classification.
It eliminates erroneous probabilities when integrating both
features for malware classification and prevents bad results from single features from 
affecting the final integrated results.

\subsection{Comparison Between API Statement Sequence Encoder and API Call Sequence Encoder}
We further compare using the API statement sequence encoder with simply using 
the API call sequence without the associated
parameters to encode the trace file's temporal characteristics using LSTM.
This is because recent methods for malware 
classification use deep learning based dynamic analysis with simple API call 
sequences, and do not consider important details such as the associated parameters.

\begin{figure}[hbt!]
\includegraphics[width=0.2\textwidth]{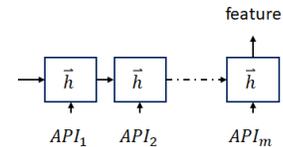}
\centering
\caption{API call sequence model. 
The first $m$ API call sequences are used for LSTM model training, after which we
take this model as a pre-trained model with which to extract temporal characteristics.} \label{APISE}
\end{figure}

For the simple API call sequence method, we use an API call sequence
model that, like most papers, uses RNN to encode the API
call sequence. We conduct experiments using LSTM to learn long-term
dependencies. The API call sequence encoder encodes temporal
series features as a fixed-length numerical vector that represents the malware behavior. 
As a feature of this method, we choose the length of the API calls in the trace file,
which determines how many API calls we evaluate in the experiment.   
Fig.~\ref{APISE} shows our API call sequence model structure. The model, which
is trained to classify malware families, is taken as a pre-trained model after training,
and called the API call sequence encoder. With it we extract API call
sequence embedding as a feature.

Shown in the table~\ref{tab12} are the training results. In this experiment, we set the sequence   
length to equal that of the API statement sequence. 
Regardless of the length, the API call sequence encoder is outperformed by
the API statement sequence encoder. This clearly shows
that considering both the API call and its associated parameters aids
in classification. This mirrors the situation in text analysis, in which
we not only take into consideration the text itself; 
for useful analysis we must also determine its actual meaning. 

\begin{table}[hbt!]
\centering
\caption{Comparison of results between API call sequence and API statement sequence.}\label{tab12}
 \begin{tabular}{|c | c | c|} 
 \hline  sequence length & Acc of API call  & Acc of API statement \\
 \hline
 100 & 44.01\% & 65.70\% \\
 \hline
 200 & 48.03\% & 67.92\% \\ 
 \hline
 300 & 50.90\% & 68.80\% \\ 
 \hline
 400 & 48.00\% &  64.74\% \\ 
 \hline
\end{tabular}
\end{table}

\section{Conclusion}
In this paper, we propose both static and dynamic analysis features for malware classification. We experiment with several ways to use the data obtained from static and dynamic analysis as features for classification. In the experimental results, we use several classification algorithms, including SVM, random forest, late fusion ensemble learning and fusion neural network. The results show that the proposed approach performs best, with an accuracy of 83.17\%. We demonstrate that when classifying malware, using integrated features works better than simply using static analysis features or dynamic analysis features, thus, providing different orientations aids in malware classification. We also show how static and dynamic features complement each other. Finally, we prove that the best malware classification performance comes when using not only the API call sequence but also its associated parameters.





%

\bibliographystyle{IEEEtran}
\bibliography{bibe.bib}
\end{document}